\documentclass[referee]{aa}
\usepackage{graphicx}
\usepackage{txfonts}  
\begin{document}

\title{Model independent approaches to reionization in the analysis of
  upcoming CMB data}

\author{Loris P.L. Colombo \and Elena Pierpaoli}

\institute{Department of Physics \& Astronomy, 
University of Southern California, Los Angeles, CA 90089-0484}
%\ead{colombo@usc.edu}

%\date{/}

\abstract{}{On large angular scales, CMB polarization depends mostly
  on the evolution of the ionization level of the IGM during
  reionization. In order to avoid biasing parameter estimates, an
  accurate and model independent approach to reionization is needed
  when analyzing high precision data, like those expected from the
  Planck experiment. In this paper we consider two recently proposed
  methods of fitting for reionization and we discuss their respective 
  advantages.}{We test both methods by performing
  a MonteCarlo Markov Chain analysis of simulated Planck data,
  assuming different fiducial reionization histories. We take into
  account both temperature and polarization data up to high multipoles,
  and we fit for both reionization and non reionization parameters.}
  {We find that while a wrong assumption on reionization may bias
  $\tau_e$, $A_s$ and $r$ by $1-3$ standard deviations, other
  parameters, in particular $n_s$, are not significantly biased. The
  additional reionization parameters introduced by considering the
  model independent methods do not affect the accuracy of the
  estimates of the main cosmological parameters, the biggest
  degradation being of order $ \sim 15\%$ for $\tau_e$. Finally, we
  show that neglecting Helium contribution in the analysis increase
  the bias on $\tau_e$, $r$ and $A_s$ even when a general fitting
  approach to reionization is assumed.}{}


\titlerunning{Model independent approaches to reionization and
 CMB data}
\authorrunning{Colombo \& Pierpaoli}
\maketitle

\section{Introduction}
\label{sec:intro}

The upcoming measurements of Cosmic Microwave Background (CMB) by the
Planck mission will allow for an unprecedented accuracy in the
determination of the CMB angular power spectra. Due to its full sky
coverage and sensitivity, Planck will provide an accurate
characterization of $E$--mode polarization autocorrelation power
spectrum, $C_l^{EE}$, at large angular scales, and either detect or
significantly improve the current limits on the $B$--mode polarization
power spectrum, $C_l^{BB}$. While other CMB polarization are currently
planned (e.g.,~\cite{taylor04,yoon06,mactavish07,samtleben08}), none
of them is expected to provide a measurement of the lowest $C_l^{EE}$
multipoles with an accuracy better than Planck. To a first
approximation, the average power of $C_l^{EE}$ on these scales depends mostly
on the optical depth to Thomson scattering due to reionization,
$\tau_e$. The value of $\tau_e$ also determines the suppression of the
intermediate to high multipoles of the temperature power spectrum,
$C_l^{TT}$. Current data by the Wilkinson Microwave Anisotropy Probe
(WMAP) imply a value $\tau_e = 0.087 \pm 0.017$, with variations of
$\Delta\tau_e \simeq 0.01$ depending on the details of the analysis
procedure and data sets considered (\cite{dunkley08}). These constraints
assume that reionization is a sharp transition occurring at a given
redshift $z_r$.
 
However, theoretical and numerical studies suggest that reionization
is a fairly complex process, possibly resulting from the sum of
several contributions occurring over different time frames
(e.g.~\cite{barkana01,venkatesan03,wyithe03,cen03,haiman03,shull07}). In
addition, observations of Ly$\alpha$ emitters in the redshift
range $6 < z < 7$, show a rapid evolution of the neutral Hydrogen
fraction of the intergalactic medium (IGM)~\cite{ota07}. In the
context of a sharp reionization, a reionization redshift $z \simeq 7$
implies $\tau_e \simeq 0.04$, and WMAP 5--year data rule out such
scenario at more than $3.5\sigma$ significance level. In order to
represent our ignorance of the reionization process, it is then
 necessary to relax the hypothesis on reionization, and consider more
complex reionization histories.

In this case, the low $C_l^{EE}$ and $C_l^{BB}$ multipoles depend not
just on $\tau_e$ but also on the detailed redshift evolution of the
(assumed homogeneous) number density of free electrons in the IGM,
$x_e(z)$, expressed in units of the Hydrogen atoms number density. For
fixed values of $\tau_e$ and all other relevant cosmological
parameters, differences in $x_e(z)$ affect the shape of the
polarization power spectra up to multipoles $l \simeq 40-50$. An
incorrect ansatz on reionization may lead to a strong bias in the
determination of $\tau_e$~(\cite{kaplinghat03,holder03,colombo05}). In
turn a bias on $\tau_e$ may result in errors on related parameter,
such as the normalization of the primordial power spectrum of density
fluctuations, $A_s$, and the tensor--to--scalar ratio $r$. At the
sensitivity level of current WMAP data, such bias is a fraction of the
experimental error, and current constraints on the optical depth can
be considered safe. In turn, this implies that constraints on the
other main cosmological parameters, in particular on $n_s$, are not
strongly dependent on the value of $\tau_e$ (\cite{dunkley08}). Planck
sensitivity, however, will be $\sim 10$ times better than WMAP 5--year
data~\footnote{Comparing the nominal single channel WMAP sensitivity
  with the specifications for Planck 143GHz channel.}  making an
accurate and model independent approach to reionization a requirement
for correct determination of $\tau_e$ and the other cosmological
parameters.

One such approach is to simply divide the redshift interval relevant
for reionization in a number of bins and try to directly constrain the
averaged value of $x_e(z)$ in each bin (\cite{lewis06}). The
implementation of the method is straightforward and allows to easily
take into account direct constraints on $x_e(z)$ (e.g. from 21cm
measurements,~\cite{tashiro08}). However, the choice of bins characteristics
is not obvious, and allowing for a fine redshift resolution implies
the addition of a significant number of strongly correlated parameters.

A principal component (PC) approach (\cite{hu03,mortonson07a}) is a
possible alternative. The reionization history is decomposed over a
set of eigenmodes, which encode the effects of a change in $x_e(z)$ on
$C_l^{EE}$. The amplitude of each eigenmode is left as a free
parameter to be determined from the data. The advantage of the method
lies in that a reduced number ($\sim 5$) eigenmodes is sufficient to
approximate the effects of a generic reionization history on the
$C_l^{EE}$'s. Using a Monte Carlo Markov Chains (MCMC) approach,
%\cite{mortonson07a,mortonson07b}
Mortonson \& Hu (2007a, 2007b) showed that PC analysis allows to
correctly recover the value of $\tau_e$, also avoiding the
introduction of spurious effects on $r$. These results considered only
the $l<100$ polarization multipoles, and assumed that the remaining
cosmological parameters were fixed to their correct value. However,
actual data analysis needs to include also temperature data and high
multipoles, and simultaneously fit for the whole set of cosmological
parameters.

CMB data allow to probe a large number of different parameters and
Planck is expected to measure the basic cosmological parameters with
high accuracy (\cite{bluebook}), providing reference values for other
kinds of measurements which probe only a subset of the parameter space
(e.g., SNIa data) and/or cover different redshift ranges and scales
(e.g., galaxy surveys, Ly$\alpha$ measurements). However, estimates of
Planck performances typically take into consideration only the basic
sharp reionization model, which can be accurately described by one
parameter. Introduction of new (reionization) parameters in the model
may give rise to new degeneracies, which in turn may bias the
estimates of the other parameters and worsen the accuracy of their
determinations. In addition, degeneracies also decrease the efficiency
of the parameter estimation procedure. In the light of the upcoming
Planck data, it is then relevant to compare how these methods affect
the whole parameter estimation process, i.e., considering also $TT$
and $TE$ spectra and high--$\ell$'s, and including also
non--reionization parameters, under the same set of conditions.

Moreover, previous studies did not take into account Helium
reionization (see, e.g., \cite{shull04,furlanetto07} and references
therein). Helium reionization has been often neglected in CMB studies,
as it contributes at most $~10\%$ of the total optical depth. However,
the Planck satellites is expected to measure $\tau_e$ with a precision
of a few percent (\cite{bluebook}) and it is interesting to study
whether Helium contribution must be explicitly accounted for in the
modeling of reionization.  In addition to the physical aspects of
reionization modeling and their impact on parameter estimation, the
computational aspects of the problem need to be factored in. The
analysis of current and future experiments require significant
numerical resources. Choosing an inappropriate parametrization can
greatly decrease the efficiency of MCMC methods, even more so when
including a large number of parameters poorly constrained by data. In
this paper we perform a comparison of the performances of the two
approaches by simulating future Planck data, corresponding to
different fiducial reionization histories both with and without Helium
contribution, and analyze them assuming either sharp reionization or
the two methods outlined below. We consider in the analysis
polarization and temperature data up to multipoles $l = 2000$, and fit
simultaneously for the main cosmological parameters. We discuss the
the advantages of each methods, both in terms of the effects on the
recovered parameters and in terms of computational cost.

The outline of the paper is as follow. In Section~\ref{sec:model}, we
briefly review the proposed model independent methods. In the
following Section~\ref{sec:analysis}, we discuss the fiducial
reionization histories considered and our simulations of experimental
data and MCMC analysis. We present our results in Section~\ref{sec:res} and
we draw our conclusions in Section~\ref{sec:conclusions}.

\section{Model Independent Approaches to Reionization}
\label{sec:model}

\subsection{Binning The Reionization History}
\label{sec:binning}
We consider the redshift set $z_0 < z_1 < z_2 < ... <z_N$,
dividing the interval $(z_0,z_N)$ into $N$ bins, so that
\begin{equation}
x_e(z) = x_{e,i}~~~~~~  z_{i -1} < z < z_i,~~i=1,...,N.
\end{equation}
In modeling the reionization history, we neglect Helium reionization
and assume $x_e(z) =1$ for $z < z_0$ while for $z > z_N$ we match
$x_e(z)$ to the small residual ionization level from incomplete
recombination. In particular, according to data on Ly$\alpha$
emitters (\cite{ota07}) and quasar spectra (\cite{fan06}), we assume $z_0
=6$. Fixing $z_N = 30$ allows for the contributions of the first stars
and/or early black holes (\cite{ricotti05}) to $\tau_e$; we ignore here
the possible X-ray emission from high--$z$ dark--matter interactions
(e.g.~\cite{hansen04,mapelli06}). The interval $(z_0,z_N)$ is then
divided into $N = 6$ equal bins.

To avoid instabilities during numerical integration, we in practice
enforce an analytical expression for $x_e(z)$:
\begin{eqnarray}
\label{eq:chi}
x_e(z) = \sum_{i=1}^{N} x_{e,i} \chi_i (z)~~~~~~  \\
\chi_i(z) = {1 \over 2} \left\{ {\rm tanh}\left[\alpha {\eta(z)-\eta(z_{i-1}) 
\over \eta(z_{i-1})}  \right] - {\rm tanh}\left[\alpha {\eta(z)-\eta(z_i) 
\over \eta(z_i)} \right] 
\right\}~; 
\end{eqnarray}
where $\eta(z)$ is the conformal time at redshift $z$ and $\alpha$
governs the sharpness of the transition. Following CAMB (\cite{camb}),
we usually take $\alpha= 150$. We also assume a flat prior on the
$x_{e,i}$. As pointed out by Lewis et al. (2006)
%~\cite{lewis06}, 
constraints on the
$x_{e,i}$ may depend significantly on the details of the binning
kernels and the priors, if the data are poor. In addition, results for
adjacent bins will usually be strongly correlated.

\subsection{Principal Component Analysis}
\label{sec:pca}
Following Mortonson \& Hu (2007a, 2007b)
%~\cite{mortonson07a,mortonson07b}, 
we divide the interval
$(z_0,z_N)$ in $N$ equal bins of width $\Delta z = 0.25$, and consider
a fiducial binned reionization history $\{x_{e,i}\}, i=1,2,...,N$. We
take $z_0 = 6$ and $ z_N =30$ and define $x_e(z)$ outside this
interval as we did in the previous section. An estimate of the
accuracy with which an experiment can measure the $\{x_{e,i}\}$ is
given by the Fisher matrix. Since we are interested in the effects of
$x_e(z)$ on CMB spectra, we can approximate the Fisher matrix as:
\begin{equation}
F_{i,j} \sim \sum_{l=2}^{l_{max}}\sum_{l'=2}^{l_{max}} {\partial C_l^{\rm EE} \over
\partial x_{e,i}}{\rm Cov}(C_l^{\rm EE},C_{l'}^{\rm EE})^{-1} 
{\partial C_{l'}^{\rm EE} \over \partial x_{e,j}}.
\end{equation}
For a full--sky noise--free experiment, the covariance ${\rm
  Cov}(C_l^{\rm EE},C_{l'}^{\rm EE}) = {2 \over 2l +1} (C_l^{\rm EE}
)^2 \delta_{ll'}$, and the main contribution to the Fisher matrix
comes from the $ l \lesssim l_{max} = 100$ multipoles of $C_l^{\rm
  EE}$. Contributions from $TT$ and $TE$ modes are negligible with
respect to those from $E$--mode polarization and we do not include
them into the definition of the Fisher matrix. In the following we
will test if this approximation is still adequate when considering
also high--$l$ $TT$ data.

The principal components of $x_e(z)$ are defined as the eigenvectors, 
$S_n(z_i)$, of the Fisher matrix:
\begin{equation}
F_{i,j} = {1 \over N^2} \sum_{n=1}^N S_n(z_i) \lambda_n^2 S_n(z_j)~;
\end{equation}
which satisfy the orthonormalization conditions:
\begin{eqnarray}
\sum_{i=1}^{N} S_n(z_i)S_m(z_j) \Delta z = (z_N -z_0) \delta_{nm}~, \\
\sum_{n=1}^{N} S_n(z_i)S_n(z_j) = N \delta_{ij}~. 
\end{eqnarray}
In the limit $\Delta z \rightarrow 0$, the first relation can be replaced
with an integral over $z$, and a generic $x_e(z)$ can be written as:
\begin{equation}
\label{eq:sum}
x_e(z) = x_e^{fid}(z) + \sum_{n=1}^N \mu_n S_n(z)~.
\end{equation}
In this representation, we replaced the $N$ values $\{ x_{e,i} \}$
defining a generic $x_e(z)$ with respect to our choice of binning,
with the $N$ mode amplitudes $\{ \mu_n \}$. Thus, in principle, the
number of parameters required to characterize a generic reionization
history has not changed. However Mortonson \& Hu (2007a)
%~\cite{mortonson07a} 
showed that most
information needed to determine CMB features is contained in the $\sim
5$ eigenmodes corresponding to the highest eigenvalues, $\lambda^2_n$,
thus allowing for a significant compression of information. When
truncating the sum in Equation~(\ref{eq:sum}), care must be taken that
the resulting $x_e (z)$ be consistent with the definition of the
Hydrogen ionization fraction, i.e. $0 < x_e(z) < 1$.

From an operative point of view, when analyzing the synthetic data, we define
$S_n(z)$ in analogy with Equation~(\ref{eq:chi}). In addition, we take
flat priors on the $\{ \mu_n \}$ and check that the resulting $x_e(z)$
does not have unphysical values.
 
\section{Analysis of Simulated Data}
\label{sec:analysis}

\subsection{Reference Models}
\label{sec:reference}
In order to test the effects on parameter estimation of the model
independent approaches discussed in the previous section, we consider
an ideal experiment with instrumental characteristics like the nominal
performance of the 143GHz Planck channel (\cite{bluebook}): Gaussian
beam of width $\theta_{\rm FWHM} = 7.1'$, temperature and polarization
sensitivities of $\sigma_{\rm T} = 42 \mu{\rm K{\cdot}arcmin}$ and
$\sigma_{\rm P} = 80 \mu{\rm K{\cdot}arcmin}$, respectively, and
assuming a sky coverage $f_{\rm sky} = 0.80$. The actual Planck
performance will exceed this specification, both due to the
availability of more frequency channels and to an actual noise level
which is better that the nominal requirements cited here.  However,
real data will require a significant foreground removal, and possibly
not all channels will be available for cosmological analysis.
While the actual Planck data analysis will therefore need to incorporate 
more subtleties, the main aim of this work is a comparison of different
approaches to reionization modeling.

\begin{figure}
\begin{center}
\includegraphics*[width=12cm]{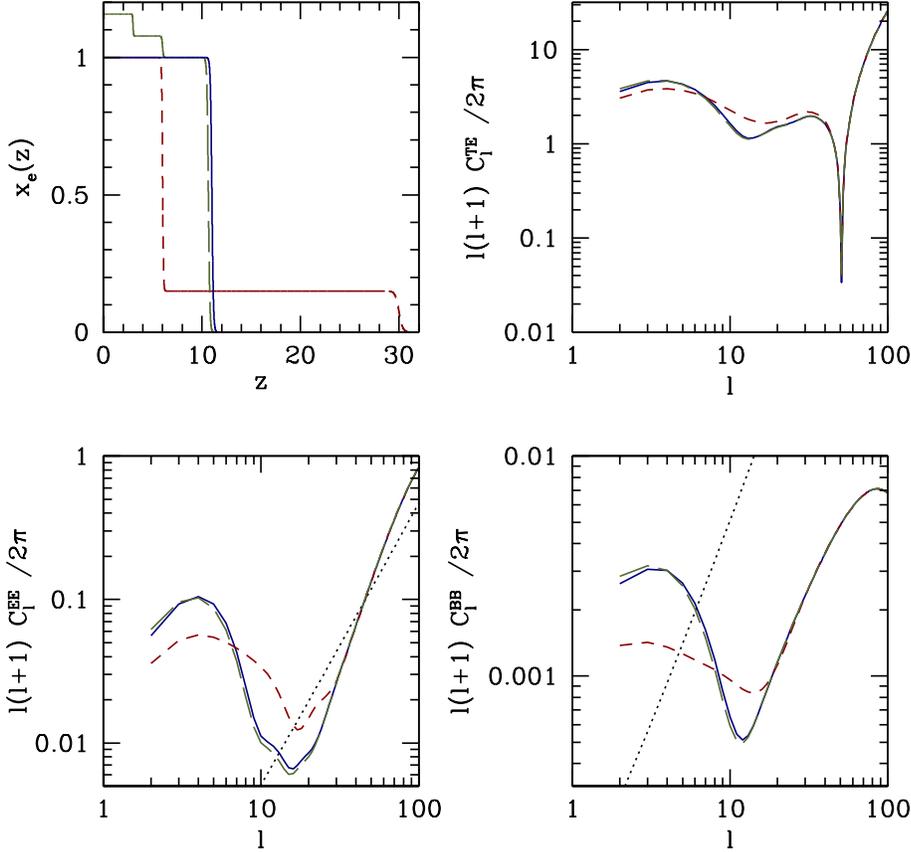}
\end{center}
\caption{\label{fig:cls} Fiducial models. Top left: $x_e(z)$ for a
  sharp reionization (solid line), two--step reionization
  (short--dashed) and Helium reionization (long--dashed). All models
  have $\tau_e = 0.085$, and the same values of the other cosmological
  parameters. The other panels show the corresponding angular power
  spectra for $TE$ (top right), $EE$ (bottom left) and $BB$ (bottom
  right). The dotted line shows the assumed Planck noise power
  spectrum.}
\end{figure}

We then generate simulated data corresponding to different fiducial
reionization histories: 1) a sharp reionization model with $\tau_e =
0.085$; 2) a model with the same $\tau_e$ but:
\begin{eqnarray}
x_e(z) & = & 1~~~~~~~~~~~ z < 6 \\
\nonumber
       & = & 0.15~~~~~~~  6 < z < 30~;
\end{eqnarray}
notice that this is the same reionization model used to define the
eigenmodes; 3) a model with the same $\tau_e$ and:
\begin{eqnarray}
x_e(z) & = & 1.158~~~~~~~~~~   z < 3 \\
\nonumber
       & = & 1.079~~~~~~~~~~   3 < z < 6 \\
\nonumber
       & = & 1~~~~~~~~~~~~~~~~   6 < z < 10.65~.
\end{eqnarray}
The values of the remaining cosmological parameters are unchanged
between the models: the physical baryon and cold dark matter densities
$\omega_b = 0.0224$ and $\omega_c = 0.112$, respectively; the slope
and amplitude of the primordial power--law spectrum of density
fluctuations $n_s = 0.95$ and ${\cal A_{\rm s}} \equiv
\log_{10}(10^{10} A_{\rm s}) = 3.135$; the Hubble parameter $H_0 = 72
$Km/s/Mpc; $r (k = 0.05 {\rm Mpc^{-1}}) = 0.1$ and $Y_{\rm He} = 0.24$
is the Helium mass fraction. The spectral index of tensor modes is
fixed according to the consistency relation for slow--roll inflation:
$n_T = -r/8 = -.0125$.  Even though the effects of reionization models
considered here are restricted to the $l <100$ (see
figure~\ref{fig:cls}) multipoles of $E$ and $B$--mode power spectra,
we take into account also temperature data, as we are interested in
assessing the effects of the different parametrizations also on
non--reionization parameters.  Each models is in turn analyzed: 1)
assuming a sharp reionization, 2) considering $N=6$ bins of $\Delta z
= 4$ between $z_0 = 6$ and $z_N = 30$, 3) using the principal
components method. Besides fitting for the reionization parameters, we
take also take $\{\omega_b,\omega_c,n_s,{\cal A_{\rm s}},H_0,r\}$ as
free parameters, while $n_T$ and $Y_{\rm He}$ are fixed to the input
values.

We notice here that complete ionization in the fiducial
sharp--reionization model happens at $z_r \simeq 11$. Thus, the
corresponding reionization history can not be correctly modeled
neither by the binning scheme we selected nor by a small (i.e $\le 5$)
number of eigenmodes. Conversely, the two--step fiducial model can be
accurately parametrized by both model independent approaches. The
third model has been chosen as a toy model of Helium
reionization. Numerical studies
suggest that Helium singly ionizes at about the same time as
Hydrogen (\cite{venkatesan03,shapiro04}), due to the closeness of the
respective ionization energies. The simplest way to account for Helium
reionization would be to assume a single reionization
event after which $x_e \simeq 1.08$. However, as discussed above,
comparison of WMAP measurements and Ly$\alpha$ observations suggest an
extended reionization process in which ionization of the IGM begins at
$z_r \sim 20$ and is completed by $z \sim 6-7$. In this kind of
scenario, Helium contribution allows $x_e$ to exceed unity only for $z
\lesssim 6$. We follow here a conservative approach and consider
doubly ionized Helium for $z < 3$ and singly ionized Helium for $3 < z
< 6$; Helium contribution to the total optical depth is then $\Delta
\tau_e \sim 0.004$.
 
\subsection{Likelihood approximation}
\label{sec:approx}
A set of CMB measurements can be represented by a set of vector
coefficients ${\bf a}_{lm} = \{ a_{lm}^T, a_{lm}^E, a_{lm}^B \}$, with
covariance matrix ${\bf C}_l \equiv \langle {\bf a}_{lm} {\bf
  a}_{l'm'}^\dagger \rangle \delta_{ll'} \delta_{mm'}$
(e.g.~\cite{percival06,hamimeche08}); the corresponding quadratic
estimator is given by:
\begin{equation} 
\label{eq:quadratic}
{\hat {\bf C}}_l = {1 \over 2l +1} \sum_{m = -l}^l {\bf a}_{lm} {\bf
  a}_{lm}^\dagger~.
\end{equation}
For a full--sky, noise--free experiment with infinite resolution, the
${\hat {\bf C}}_l$ are distributed according to a Wishart
distribution. Using Bayes theorem, the corresponding log--likelihood,
normalized so that ${\cal L} = 1$ for ${\bf C}_l ={\hat {\bf C}}_l $,
is given by:
\begin{equation}
\label{eq:likelihood}
-2 \ln {\cal L} (\{{\bf C}_l \}| \{{\hat {\bf C}}_l\}) =
\sum_{l=2}^{l_{max}} (2l +1) \left [ {\rm Tr} ({\hat {\bf C}}_l {\bf
    C}_l^{-1}) -\ln (|{\hat {\bf C}}_l{\bf C}_l^{-1}|) -3 \right]~.
\end{equation}
In the presence of white isotropic noise and assuming a perfect
Gaussian beam, the above expression is still valid if we replace
${\bf C}_l$ with ${\bf C}_l +{\bf N}_l$, where ${\bf N}_l = {\rm diag}
({\cal N}_l^T,{\cal N}_l^P,{\cal N}_l^P)$ is the noise correlation
matrix, while the noise power spectrum ${\cal N}_l^T = \sigma_{\rm
  T}^2 \exp \left[ l(l+1) \theta_{\rm FWHM}^2 /(8 \ln 2)\right]$ and
similarly for polarization. An analogous substitution is required for
the estimator ${\hat {\bf C}}_l$. 

If full--sky measurements are not available, the spherical harmonics
coefficients for the cut sky, ${\tilde a}_{lm}^X$, are a linear
combination of true spherical harmonics coefficients corresponding to
different modes and multipoles,
\begin{equation}
\label{eq:atilde}
{\tilde a}_{lm}^X = \sum_Y \sum_{l'm'} K^{XY}_{lm l'm'} a_{l'm'}^Y
\end{equation}
where the kernels $K^{XY}_{lm l'm'}$ encodes the effect of non uniform
sky coverage~\cite{hivon02}. In this case
Equation~(\ref{eq:likelihood}) is no longer valid, although for
azimuthal symmetric cuts an analytic expression for the likelihood
can still be evaluated (\cite{lewis02}), although at the cost of a
significantly increased computational time. Here, instead, we suppose
that the mode coupling resulting from incomplete sky coverage can be
accounted for by multiplying Equation~(\ref{eq:likelihood}) by a
factor $f_{sky}^2$. Although this approximation does not correctly
account for mode mixing, in particular $E$--$B$ mixing, in this way
the likelihood functions still peaks at the full sky value, therefore
any bias we find in our results is due to the modeling of reionization
rather than the likelihood approximation. In addition, while errors on
parameters may not be correctly estimated by this approximation, a
correct assessment of errors would need to take into account the
actual details of the data analysis pipeline, including tod filtering,
map making and foreground removal, which are beyond the scope of this
paper.

We then perform a MCMC analysis of the simulated data
using a version of the CosmoMC package (\cite{cosmomc}) modified to take
into account different reionization models. We also fix $l_{max} =
2000$.  We determine convergence of our chains by requiring that the
Gelman \& Rubin ratio be $R -1 < 0.05$ and simultaneously checking the
stability of the $95\%$ confidence limit on all parameters. In
practice, this latter criterion leads to $R -1 \sim 0.02-0.03$ for the
converged chains.

\section{Results}
\label{sec:res}

\begin{figure}
\begin{center}
\includegraphics*[width=12cm]{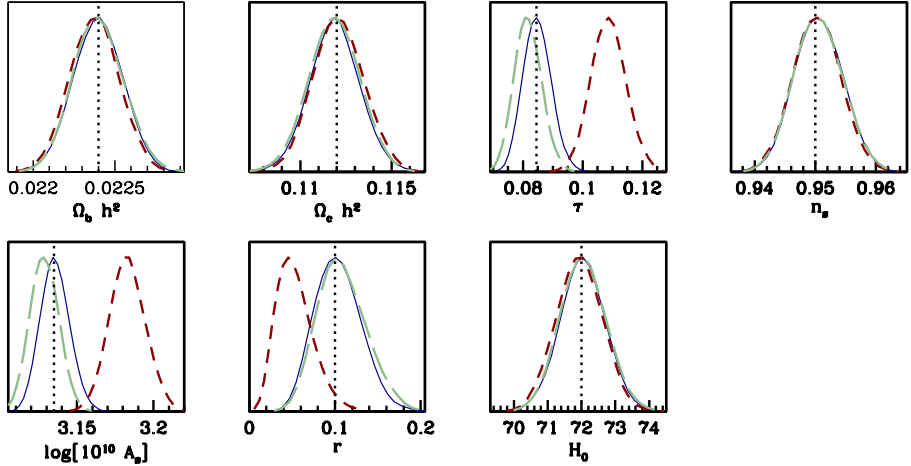}
\end{center}
\caption{\label{fig:confr1} Results for MCMC analysis assuming a sharp
  reionization for three different fiducial models: sharp reionization
  (solid lines), two--step reionization model (short--dashed) and
  Helium reionization (long--dashed). The reference value of the input
  parameters, shown by the vertical dotted lines, are the same in all
  fiducial models}
\end{figure}

First of all, in order to identify how parameters can be affected by
an incorrect assumption on reionization, we can compare the results of
analyzing the three reference models assuming a sharp reionization.
In Figure~\ref{fig:confr1} we show the resulting marginalized
distribution. When an incorrect assumption on reionization is made the
estimate of $\tau_e$ is biased by 1 or more standard deviations,
depending on the fiducial reionization history, in agreement with
previous
findings (\cite{kaplinghat03,holder03,colombo05,mortonson07a}). For the
models considered here, the bias is more relevant in the case of the
two--step fiducial model, as in this scenario reionization starts
significantly earlier than in either of the other models considered;
thus the range of multipoles affected is greater. The corresponding
numerical values are summarized by the third columns of
tables~\ref{tab:twostep} through~\ref{tab:helium} for the sharp,
two--step and Helium reionization fiducial models described in
Section~\ref{sec:reference}, respectively. For each parameter, we
report the mean estimated by the chains.

%Table~\ref{tab:twostep} through~\ref{tab:helium} report the results of
%the MCMC analysis for the sharp, two--step and Helium reionization
%fiducial models described in section~\ref{sec:reference},
%respectively. Each fiducial model has been analyzed with different
%assumptions on reionization and bold faced entries show when the estimate
%of a parameter differ from the input value by more than half standard
%deviation. Several considerations can be made. 

\begin{table}[!hbt]
\begin{center}
\begin{tabular}{cccccc}
\hline \hline
\multicolumn{4}{c}{{\rm Two--step reionization}} \\
\hline \hline
& fiducial & SR  &  BR & E3 &E5  \\
& value    &    & & & \\
$100 \omega_b  $& $2.24$ &$ 2.238 \pm 0.015$&$2.236 \pm 0.015$&
             $2.238 \pm 0.015$&  $2.235 \pm 0.015$ \\
$\omega_c  $&$.112$&$.1121  \pm .0013 $&$.1120  \pm .0014 $&
             $ .1121  \pm .0013$& $ .1120  \pm .0014$ \\
$\tau_e      $&$.085$&${\bf .1087  \pm .0057}$&$.0874  \pm .0055 $&
             $ .0862  \pm .0051$& $ .0877  \pm .0056$ \\
$n_s       $&$.950$&$.9502  \pm .0038 $&$.9496  \pm .0039 $&
             $ .9492  \pm .0038$&$ .9495  \pm .0041$ \\
${\cal A}_s$&$3.135$&${\bf 3.183   \pm 0.011 }$&$3.140   \pm 0.011  $&
             $  3.138   \pm 0.010$&$3.138   \pm 0.011$ \\
$r         $&$.10$&${\bf .053   \pm .021 }$&$.096   \pm .027  $&
             $  .100   \pm .027$&$.098   \pm .027  $ \\
$H_0       $&$72$&$71.94   \pm 0.66   $&$71.93   \pm 0.67   $&
             $  71.88   \pm 0.66 $&$71.92   \pm 0.68   $\\
\hline\hline
\end{tabular}

\caption{\textsl{Parameter estimates for a two--step reionization
    fiducial model with $\tau_e = 0.085$ and other parameters as
    specified in the text, assuming: a sharp reionization (SR), a
    binned reionization (BR), either 3 (E3) or 5 (E5) principal
    components eigenmodes. Bold faced entries show when the bias
    between the input and recovered value exceeds half the associated
    error.}}
\label{tab:twostep}
\end{center}
\end{table}

\begin{table}[!hbt]
\begin{center}
\begin{tabular}{ccccc}
\hline \hline
\multicolumn{4}{c}{{\rm Sharp reionization}} \\
\hline \hline
&fiducial & SR  &  BR & E5  \\
&value    &     &     &\\
$100 \omega_b  $&$.224$&$ 2.240 \pm 0.015$&$2.242 \pm 0.015$&
             $2.242 \pm 0.015$ \\
$\omega_c  $&$.112$&$.1119  \pm .0013 $&$.1118  \pm .0013 $&
             $ .1117  \pm .0013$ \\
$\tau_e      $&$.085$&$.0848  \pm .0046 $&$.0853  \pm .0053 $&
             $ .0830  \pm .0052$ \\
$n_s       $&$.950$&$.9506  \pm .0039 $&$.9509  \pm .0039 $&
             $ .9505  \pm .0040$ \\
${\cal A}_s$&$3.135$&$3.136   \pm 0.009  $&$3.137   \pm 0.010  $&
             $  3.132   \pm 0.009$ \\
$r         $&$.10$&$.104   \pm .026  $&${\bf .118   \pm .027} $&
             ${\bf  .120   \pm .029}$ \\
$H_0       $&$72$&$72.03   \pm 0.66   $&$72.14   \pm 0.66   $&
             $  72.12   \pm 0.63 $ \\
\hline\hline
\end{tabular}

\caption{\textsl{ Same as Table~\ref{tab:twostep} but for a sharp
    reionization fiducial model analyzed assuming: a sharp
    reionization (SR), a binned reionization (BR), 5 (E5) principal
    components eigenmodes. }}
\label{tab:sharp}
\end{center}
\end{table}

\begin{table}[!hbt]
\begin{center}
\begin{tabular}{ccccc}
\hline \hline
\multicolumn{4}{c}{{\rm Helium reionization}} \\
\hline \hline
&fiducial & SR  &  BR & E5  \\
&value    &     &     &     \\ 
$100 \omega_b $&$2.24$&$ 2.240 \pm 0.015 $&$ 2.242 \pm 0.015$  &$ 2.242 \pm 0.014$
              \\
$\omega_c     $&$.112$&$.1119 \pm .0013 $&$.1116 \pm .0013$& $.1117 \pm .0013 $
              \\
$\tau_e       $&$.085$&${\bf .0816 \pm .0044}$&$.0831 \pm .0049 $&${\bf .0820  \pm .0047}$
              \\
$n_s          $&$.95$&$.9506  \pm .0039$&$.9512  \pm .0039$&$.9508  \pm .0039 $
              \\
${\cal A}_s   $&$3.135$&${\bf 3.129   \pm 0.009 }$&$3.132   \pm 0.010 $&$3.132   \pm 0.009  $
              \\
$r            $&$.10$&$.108   \pm .028  $& ${\bf .124   \pm .027 }$ &${\bf .123   \pm .027}$
              \\
$H_0          $&$72$&$72.05   \pm 0.68   $& $72.18   \pm 0.65 $&$72.13   \pm 0.64   $
              \\
\hline\hline
\end{tabular}

\caption{\textsl{Same as Table~\ref{tab:twostep} but for a Helium
    reionization fiducial model analyzed assuming: a sharp
    reionization (SR), a binned reionization (BR), 5 (E5) principal
    components eigenmodes.}}
\label{tab:helium}
\end{center}
\end{table}

In turn, an incorrect determination of $\tau_e$ biases the value of
the amplitude of the primordial power spectrum according to $A_s
e^{-2\tau_e} = const$, and also results in a wrong determination of
$r$, as pointed out by \cite{mortonson07b}. Notice that here we fixed
the value of the tensor spectral index; leaving $n_T$ as a free
parameter would slightly increase the error on $r$, reducing the
significance of the discrepancy. Other parameters are mostly
unaffected. In particular, it is interesting to notice that the
distribution for $n_s$ does not depend on the reionization priors,
even though information from polarization is critical to break the
$\tau_e$--$n_s$ degeneracy present in $TT$ spectra. This is due to the
fact that at the sensitivity level considered here it is possible to
get information on $n_s$ from the $l>100$ multipoles; on these scales
the $n_s$ affect the $l$--scaling of the power spectrum, while the
reionization history affect the $C_l$'s with an overall suppression
depending on the value of $\tau_e$. With Planck $l$--leverage, it is
possible to disentangle the effects of $n_s$ and $\tau_e$, and recast any
uncertainty on $\tau_e$ on the value of $A_s$. Since deviation of $n_s$ from
unity allow to place constraints on the shape of inflation potential,
we can conclude that such constraints will be safe even if an
incorrect assumption on reionization is made.

\begin{figure}
\begin{center}
\includegraphics*[width=12cm]{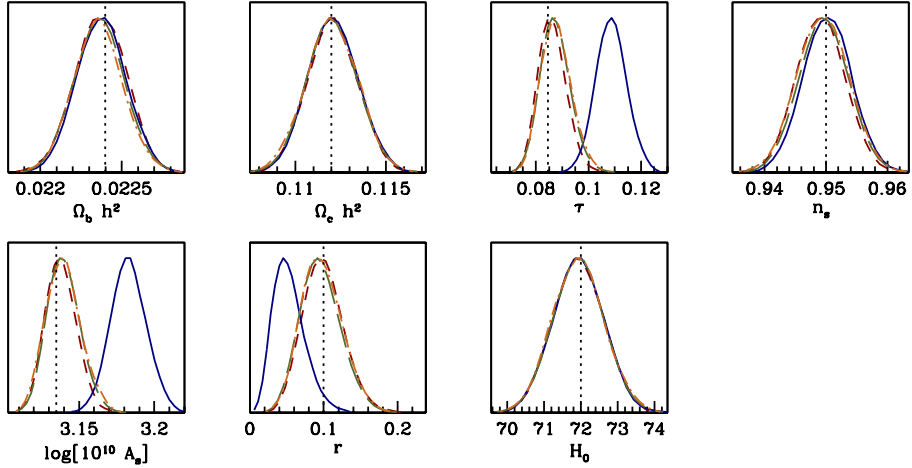}
\end{center}
\caption{\label{fig:confr2} Results for MCMC analysis for two--step
  reionization fiducial model. Model has been analysed assuming a
  sharp reionization (solid lines), a binned reionization
  (long--dashed) and using 3 (short--dashed) or 5 (dot--dashed)
  principal components.}
\end{figure}

\begin{figure}
\begin{center}
\includegraphics*[width=12cm]{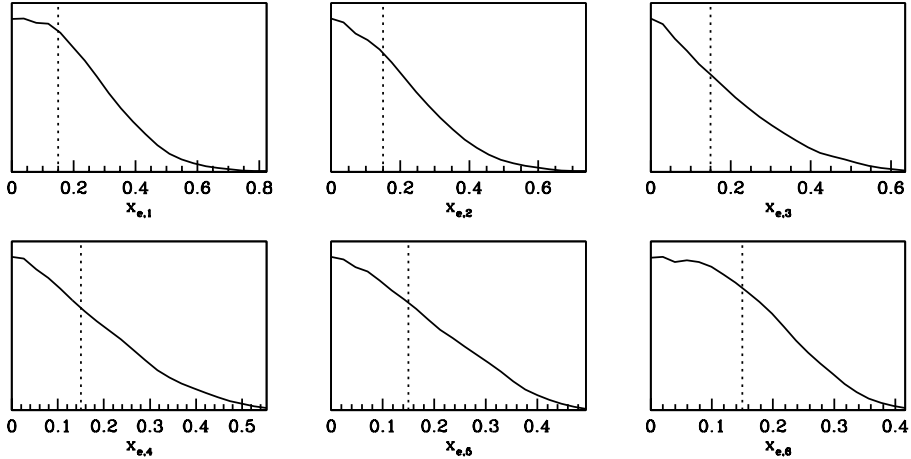}
\end{center}
\caption{\label{fig:xe_step} Constraints on values of $x_e(z)$ in bins
  of width $\Delta z = 4$ between $z_0 = 6$ and $z_N = 30$ for a
  two--step reionization fiducial model. The dashed line show the
  input value of $x_e(z)$ in each bin.}
\end{figure}

We next turn to the model independent
approaches (\cite{lewis06,mortonson07a,mortonson07b}). In particular, we
assess the effect of introducing these additional parameters on the
whole analysis procedure, on the same set of data. As a first case, we
consider a fiducial two--step reionization model. In principle,
fitting this reionization would require a low number of reionization
parameter, e.g. a single value of $x_e(z)$ for $6 < z < 30$, or a
single eigenmode. Here instead, we run MCMC assuming a $N=6$ redshift
bins, or either 3 or 5 eigenmodes. By considering more reionization
parameters than effectively needed, we can study whether a bias is
introduced or if error estimates are affected.

Figure~\ref{fig:confr2} shows the resulting marginalized
likelihoods. For reference purpose, we also repeat results for the
sharp reionization history. Notice that for the binned and the PC
analysis, $\tau_e$ is a derived parameter.  It is clear that in this
case the modeling of reionization does not bear a strong impact on the
estimates of the various parameters.  Estimates of $\tau_e$ shift by
$\sim 0.2\sigma$ depending on the methods considered, while the
corresponding error for the 3 eigenmodes analysis is $\sim 10\%$
smaller than in the other cases. There are no other significant
differences between the $3$ and $5$ eigenmodes, or between eigenmodes
and bins results. Both the binned reionization and principal
components methods slightly overestimate $\tau_e$, however the
difference between input and recovered values are below half a
standard deviation and are compatible with the statistical
uncertainty. In fact, the Gelman and Rubin convergence diagnostic $R$
roughly translates into $R \simeq 1 +r_x$, where $r_x$ is the ratio
between the variance of the sample mean and the variance of the target
distribution (\cite{dunkley05}), so that a value $R -1 \sim 0.05$
corresponds to an uncertainty on the mean of about 25\% of the
measured standard deviation.

Reconstruction of the reionization history, on the other hand, is not
particularly accurate.  Figure~\ref{fig:xe_step} shows constraints on
the value of the ionization fraction $x_{e,i}$ in the different
bins. Only weak upper limits are found: the target model could be
clearly described by just a single value of ${x_e}$ and the data do
not allow to significantly constrain the additional
parameters. However, it is reassuring that this does not have any
adverse effect on the accuracy with which data can constrain $\tau_e$
and other relevant parameters. This is possibly because, while $x_e$
bins are highly correlated (\cite{lewis06}), we do not find significant
degeneracies between the reionization parameters and the remaining
cosmological ones. A qualitatively similar conclusion holds for the
eigenvalue method: even though we add a significant number of poorly
constrained parameters, the accuracy on the reconstruction of the main
cosmological parameters is unaffected. In general, we find that the
different model independent approaches considered lead to shift in the
estimates of $0.1-0.2\sigma$; in the case of non--reionization
parameters, this holds also for the sharp $\tau_e$ analysis. Error
estimates increase at most by $10-15\%$ over the sharp reionization
value.

\begin{figure}
\begin{center}
\includegraphics*[width=12cm]{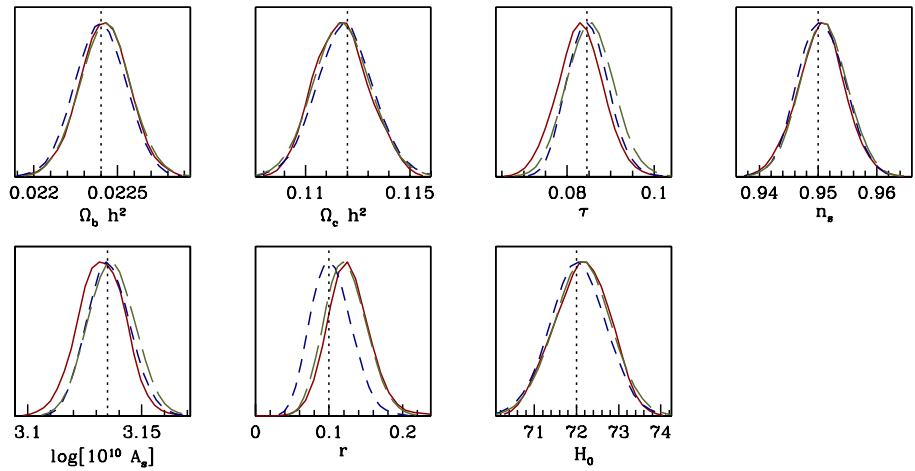}
\end{center}
\caption{\label{fig:confr3} Results for MCMC analysis for a sharp
  reionization fiducial model. Model has been analysed assuming a
  sharp reionization (solid lines), a binned reionization
  (long--dashed) and using 5 principal components (short--dashed).}
\end{figure}

As a second case, we consider a fiducial sharp reionization model,
analysed with either $N=6$ redshift bins or 5 PC eigenmodes
(see Figure~\ref{fig:confr3}). This case is conceptually opposite to the
previous one, as the fiducial reionization history cannot be
accurately described by either modeling we considered. In this case,
we do not consider the 3 PC model, as in general 3 eigenmodes are not
enough to accurately recover the reionization parameters even when all
other parameters are kept fixed (\cite{mortonson07a}).

Also in this case, both parametrizations considered measure $\tau_e$
without any relevant bias, although the 5 PC method slightly
underestimate it. Again, the discrepancy is compatible with the
statistical uncertainty. When either model independent approach is
assumed, the error on $\tau_e$ increase by just $\sim 15\%$ over the
error that would be obtained by analyzing the data assuming a sharp
reionization. It is interesting to note that both modeling
overestimate $r$ by $\sim 0.6 -0.7$ standard deviations. While this
does not represent a significant bias, it could be an hint that the
modeling of reionization considered is not fully adequate to the
underlying data and more eigenmodes, or a different binning scheme,
need to be included in the analysis. In general, checking that results
are consistent between different parametrizations allow to minimize
these spurious effect.

\begin{figure}
\begin{center}
\includegraphics*[width=12cm]{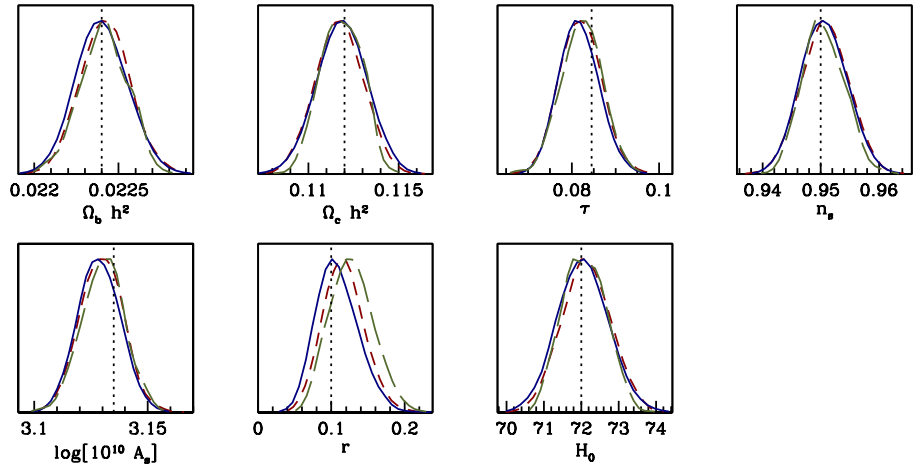}
\end{center}
\caption{\label{fig:confr4} Results for MCMC analysis for the Helium
  reionization fiducial model. Model has been analysed assuming a
  sharp reionization (solid lines), a binned reionization
  (long--dashed) and using 5 principal components (short--dashed).}
\end{figure}

We next consider the impact of Helium reionization.  Let us recall
that both the bins and PC approaches we implemented here assume that
$x_e(z) =1$ at $z \le 6$, therefore Helium reionization is not
accounted for in the modeling using for data analysis. Comparing
results for this case with those for the sharp reionization fiducial
history, then, allows to establish whether Helium reionization needs
to be included in the modeling. Results of the analysis are
summarized in Table~\ref{tab:helium} and Figure~\ref{fig:confr4}.

At the sensitivity level considered, we find that the estimated value
of $\tau_e$ is consistent with the fiducial value at the $1\sigma$
level, regardless of the assumptions on reionization. More in detail,
assuming a sharp reionization or using 5 PC, $\tau_e$ is biased by $0.6
-0.8\sigma$, while using bins $\tau_e$ is recovered within $0.5\sigma$
from the input value. In addition, $r$ is overestimated by $\sim
0.9\sigma$, both using bins and PC, while assuming sharp reionization
$r$ is recovered without a significant bias. This is due to the fact
that, for fixed $\tau_e$, Helium contribution alters the reionization
history at $z \lesssim 6$ and therefore increases the power in $EE$
and $BB$ spectra at multipoles $l <5$, with respect to a sharp
reionization model. On the other hand, in extended reionization
scenarios, power is shifted from $l <5$ multipoles to $10< l <30 $
multipoles, as reionization starts earlier than in a sharp
reionization scenario with the same $\tau_e$
(\cite{kaplinghat03,colombo05}). For the sharp reionization fiducial
model, instead, the model independent approaches overestimated $r$ by
$\sim 0.6 -0.7\sigma$ and $\tau_e$ was recovered to within half a
standard deviation, regardless of the assumptions on reionization (see
Table~\ref{tab:sharp}). This suggests that Helium reionization must
be explicitly taken into account in our modeling, more so if we
consider that the actual Planck performance is likely to exceed the
conservative specifications assumed here, and Helium contribution is
probably higher than that of our conservative approach. However, a
more in depth study of the impact of Helium reionization in CMB data
analysis is required.

Finally, we briefly discuss the computational costs of the different
approaches. For the fiducial histories considered, we found that
chains assuming 6 $x_e$ bins take 30-50\% more time to converge than
those assuming 5 PC eigenmodes. In principle, under ideal condition,
MCMC methods scale linearly with the number of parameters, so we can
expect the chains for the binned analysis to take $\sim 10\%$ more,
simply due to the different number of parameters in the two
models. However, reaching this theoretical limit is significantly
dependent on an efficient proposal distribution, i.e. on an accurate
covariance matrix in the case of Gaussian proposal
densities (\cite{dunkley05}). In this work, for each model we run a
preliminary set of chains of 60000 points (total) to determine a
starting covariance matrix. The difference in convergence times we
found here suggests that the orthogonality of the eigenmodes allow for
a more efficient exploration of the parameter space, even though such
orthogonality holds properly only when the target reionization history
is near the fiducial model used to define the
eigenmodes (\cite{mortonson07a}).  A full assessment of this point
would, however, require more simulations than those performed in this
work, and is likely to depend on the details of the actual
reionization history assumed as a fiducial model.

\section{Conclusions}
\label{sec:conclusions}

With the advent of high precision CMB polarization measurements, a
detailed modeling of reionization becomes of great relevance, both to
better constrain the detail of reionization itself and to avoid biases
on the cosmological parameters, in particular those related to
inflation. If the actual reionization history is not a single, quick
transition, assuming a sharp reionization while analyzing data may
lead to bias of 1 or more standard deviations on parameters like $\tau_e$,
$A_s$ and $r$. However, a full theoretical understanding of
reionization is still lacking and consensus on a physically motivated
parametrization of the effects of reionization on CMB has not aroused
yet. Thus, two model independent parametrizations have been recently
proposed: using a binned reionization history (\cite{lewis06})and
principal component approach (\cite{hu03,mortonson07a}).

In this work we have considered both approaches and applied them to
simulated Planck data in order to asses their accuracy and to find out
any side--effect on the estimation of the other cosmological
parameters. We considered fiducial models with the same values of all
cosmological parameters, but with different reionization histories, and
we analyzed these models assuming a sharp reionization or using both model
independent approaches. In our analysis, we included $TT$, $TE$, $EE$
and $BB$ spectra up to multipoles $l_{max} = 2000$, and we fitted both
for the reionization parameters and for the remaining cosmological parameters.
In agreement with previous results considering only $EE$ data and/or low
multipoles (\cite{kaplinghat03,colombo05,mortonson07a}), we found that the
sharp reionization analysis give accurate results only when the
fiducial model is not significantly different from a sharp
reionization history, while in general biases of order $1$--$3$
standard deviations can be expected on $\tau_e$, $A_s$ and $r$.

On the other hand, we found that both model independent methods are
able to correctly recover the various parameters; none of the
approaches provide a significant advantage over the other in term of
accuracy of the recovered parameters. More in detail, the correct
value of $\tau_e$ and $A_s$ are recovered to better than half a
standard deviation. The additional parameters, either for bin
reionization or for principal components, increase the error in
$\tau_e$ by $\sim 15\%$, but do not affect the error on the other
parameters. However, when the target model is not accurately described
by the adopted parametrization, we noticed a residual bias on $r$, of
order $0.6-0.7\sigma$. While this level of bias can be considered
safe, it nonetheless indicates that our parametrization can be
refined.  More in general, to further reduce this bias, it is
helpful to include as much external information on reionization than
can be available, e.g. using 21 cm measurements.  These external
constraints can be directly implemented into a binned reionization
approach, while for PC analysis additional work is required. 

It is worth noticing that estimates of the remaining parameters, such
as $\omega_b$, $\omega_c$ and $n_s$, are largely unaffected by the
assumptions on reionization. This is valid not only when a model
independent description of reionization is adopted, but also when the
modeling of reionization assumed for data analysis is not an adequate
description of the actual reionization history. In particular,
estimates of $n_s$ from current and future experiments can be
considered safe, regardless of the details of reionization.

We also considered a toy model of Helium reionization, in order to
assess whether such contribution must be explicitly accounted for. We
assumed a fiducial model with Helium contributing to the ionization
fraction at redshifts $z <6$, however both the bin reionization and
the principal component method used for analyzing the simulated data
assumed $x_e(z) =1$ for $z \le 6$. A comparison of the results of this
analysis, with those for a sharp reionization fiducial model with the
same $\tau_e$, show that the discrepancy between fiducial and
recovered value increase by $\sim 0.2-0.5\sigma$ for $\tau_e$, $A_s$ and
$r$, in the case of the Helium reionization fiducial model.  However,
in no case we detected a bias of $1\sigma$ or more. Even though
Helium contributes to the total optical depth for $\Delta \tau_e =
0.004$, compared to an expected error $\sigma(\tau_e) \simeq 0.005$,
these results suggest that Helium reionization need to be explicitly
taken into account in the analysis of future data.

Finally, we point out that, while we did not found significant
differences between the model independent approaches considered in
terms of the accuracy of the recovered parameters, in order to reach
convergence a bin reionization approach requires $\sim 30-50\%$ more
time than a principal component method. Since the analysis of
current and future CMB data sets require significant computational
resources, this latter aspect needs to be taken into account when
choosing a modeling of reionization.

\begin{acknowledgements} 
This work is supported by NASA grant NNX07AH59G and JPL--Planck
subcontract no. 1290790. EP is an NSF--ADVANCE fellow (AST--0649899)
also supported by JPL SURP award no. 1314616. We acknowledge the use
of the CAMB and CosmoMC packages, as well as the computational
resources provided by the University of Southern California's Center
for High-Performance Computing and Communications. The authors thank
Caltech for hospitality. 
\end{acknowledgements}


\begin{thebibliography}{99}

\bibitem[Barkana \& Loeb 2001]{barkana01} Barkana, R. \& Loeb, A. 2001,
  \prd, 349, 125--238


\bibitem[http://www.camb.info/]{camb} CAMB: http://www.camb.info/

\bibitem[Cen 2003]{cen03} Cen, R. 2003, \apj, 591, 12--37

\bibitem[Colombo et al. 2005]{colombo05} Colombo, L.P.L. et al. 2005,
  \aap ,  435, 413--420

\bibitem[http://cosmologist.info/cosmomc/]{cosmomc} CosmoMC
  http://cosmologist.info/cosmomc/

\bibitem[Dunkley et al. 2005]{dunkley05}Dunkley, J. et al. 2005,
  \mnras, 356, 925-936

\bibitem[Dunkley et al.2008]{dunkley08} Dunkley J et al. 2008
  [preprint arXiv:0803.0586v1]

\bibitem[Fan et al. 2006]{fan06} Fan, X. et al. 2006, \aj,  132, 117

\bibitem[Furlanetto \& Peng 2007]{furlanetto07} Furlanetto, S.R. \&
  Peng, O.S. 2007, \aj  submitted [preprint
    arXiv:0711.1542v1]

\bibitem[Gnedin \& Fan 2006]{gnedin06} Gnedin, N.Y. \& Fan, X. 2006,  
\apj , 648,  1

\bibitem[Haiman \& Zolder 2003]{haiman03} Haiman, Z. \& Holder, G.P. 2003,
\apj, 595, 1--12

\bibitem[Hamimeche \& Lewis 2008]{hamimeche08} Hamimeche,S. \& Lewis, A.
  2008 [preprint arXiv:0801.0554v1]


\bibitem[Hansen \& Haiman 2004]{hansen04} Hansen, S.H. \& Haiman, Z.
  2004, \apj,  600, 26-31

\bibitem[Hivon et al. 2002]{hivon02} Hivon, E. et al. 2002, \apj, 
 567, 2

\bibitem[Holder et al. 2003]{holder03} Holder, G. et al. 2003, \apj,
 595, 13--18

\bibitem[Hu \& Holder 2003]{hu03} Hu, W. \& Holder, G.P. 2003, \prd,
 68, 023001

\bibitem[Kaplinghat et al. 2003]{kaplinghat03} Kaplinghat, M. et
  al. 2003, \apj, 583, 24--32

\bibitem[Lewis et al. 2002]{lewis02} Lewis, A. et al. 2002, \prd,
  65, 023505

\bibitem[Lewis et al. 2006]{lewis06} Lewis A., Weller, J. \& Battye, R.
  2006, \mnras, 373, 561--570

\bibitem[MacTavish et al. 2007]{mactavish07} MacTavish, C.J. et al. 2007
  [preprint arXiv:0710.0375v1]

\bibitem[Mapelli et al. 2006]{mapelli06} Mapelli, M. et al. 2006, \mnras,
 369, 1719-1724

\bibitem[Mortonson \& Hu 2007a]{mortonson07a} Mortonson, M.J. \& Hu, W.
  2007, \apj   submitted [preprint arXiv:0705.1132v1]

\bibitem[Mortonson \& Hu 2007b]{mortonson07b} Mortonson, M.J. \& Hu, W.
  2007, \apj  submitted [preprint arXiv:0710.4162v1]

\bibitem[Ota et al. 2007]{ota07} Ota, K. et al. 2007, \apj, [preprint
  arXiv:0707.1561v1]

\bibitem[Percival \& Brown 2006]{percival06} Percival, W.J. \& Brown, M.
  L. 2006, \mnras, 372, 1104

\bibitem[Planck Blue Book]{bluebook} Planck Blue Book: 
http://www.rssd.esa.int/SA/PLANCK/docs/Bluebook-ESA-SCI(2005)1\_V2.pdf

\bibitem[Ricotti et al. 2005]{ricotti05} Ricotti, M., Ostriker, J.P. \&
  Gnedin, N.Y. 2005, \mnras, 357, 207

\bibitem[Samtleben 2008]{samtleben08} Samtleben, D. 2008 in the
  Proceedings `A Century of Cosmology', San Servolo (Venezia, Italy),
  August 2007 [preprint arXiv:0802.2657v1]

\bibitem[Shapiro et al. 2004]{shapiro04} Shapiro, P.R. et al. 2004,
\mnras,  348, 753--782

\bibitem[Shull et al. 2004]{shull04} Shull, J.M. et al. 2004, \apj, 600, 570

\bibitem[Shull \& Venkatesan 2007]{shull07} Shull, J.M. \& Venkatesan,A.
  2007, \apj, submitted [preprint
    arXiv:astro-ph/0702323v1]

\bibitem[Tashiro et al. 2008]{tashiro08} Tashiro, H. et al. 2008,
\mnras, submitted [preprint arXiv:0802.3893v1]

\bibitem[Taylor et al. 2004]{taylor04} Taylor, A. et al. 2004 in
  roceedings of the XXXVIXth Rencontres de Moriond "Exploring the
  Universe" [preprint arXiv:astro-ph/0407148v1]

\bibitem[Venkatesan et al. 2003]{venkatesan03} Venkatesan, A., Tumlinson,
  J. \& Shull, J.M. 2003, \apj, 584, 621--632

\bibitem[Wyithe \& Loeb 2003]{wyithe03} Wyithe, S. \& Loeb, A. 2003, \apj,
  586, 693--708

\bibitem[Yoon et al. 2006]{yoon06} Yoon, K.W. et al. 2006 in Millimeter
  and Submillimeter Detectors and Instrumentation for Astronomy III,
  Proceedings of SPIE, 6275, 2006 [preprint arXiv:astro-ph/0606278v1]





\end{thebibliography}
\end{document}